\documentclass[a4paper,11pt]{article}
\pdfoutput=1 

\usepackage{jinstpub} 

\newcommand{\MAPbBr}{L$_2[$MAPbBr$_3]$PbBr$_4$ } 
\newcommand{\FAPbBr}{L$_2[$FAPbBr$_3]$PbBr$_4$ } 
\newcommand{\MAPbI}{L$_2[$MAPbI$_3]$PbI$_4$ } 
\newcommand{\FAPbI}{L$_2[$FAPbI$_3]$PbI$_4$ } 
\newcommand{\CsPbI}{L$_2[$CsPbI$_3]$PbI$_4$ }

\title{Light Yield of Perovskite Nanocrystal-Doped Liquid Scintillator}

\author[a,b,1]{D. Gooding,\note{Corresponding author.}}
\author[b]{J. Gruszko,}
\author[a]{C. Grant,}
\author[c]{B. Naranjo,}
\author[b]{and L. Winslow}

\affiliation[a]{Boston University, Department of Physics, \\590 Commonwealth Avenue
Boston, MA 02215, USA}
\affiliation[b]{Massachusetts Institute of Technology, Department of Physics and Laboratory for Nuclear Science,\\77 Massachusetts Ave Cambridge, MA 02139, USA}
\affiliation[c]{University of California Los Angeles, Department of Physics \& Astronomy,\\475 Portola Plaza, Los Angeles, CA 90095-1547, USA}

\emailAdd{dgooding@bu.edu}

\abstract{Liquid scintillators doped with metals are needed for a variety of measurements in nuclear and particle physics. Nanoparticles provide a mechanism to dope the scintillator and their unique optical properties could be used to enhance detection capabilities. We present here the first study of lead-based perovskite nanoparticles for this application. Perovskites are an attractive choice due to the versatility of their crystal structure and their ease of synthesis.}


\begin{document}
\maketitle
\flushbottom

\section{Introduction}
\label{sec:intro}

Liquid scintillator-based detectors have been at the center of many of the great discoveries of neutrino physics, from the first detection of neutrinos, to the discovery of neutrino oscillations, to current precision measurements with experiments such as KamLAND~\cite{kamReactor2013}, Borexino~\cite{borexinoSoloar2011}, Daya Bay~\cite{dayabay2016}, RENO~\cite{reno2018s}, Double Chooz~\cite{doublechooz2014} and PROSPECT~\cite{prospect2018}. The technology is well-suited to neutrino physics because it provides cost-effective scaling to kiloton-scale masses while providing straight-forward background suppression through purification, spatial and temporal coincidence analyses, self-shielding, and pulse shape discrimination. While the achievements of current liquid scintillator experiments are impressive, the next generation of measurements has even more stringent background requirements and requires higher concentrations of dopants with a more varied set of isotopes. 

Liquid scintillator detectors are large calorimeters. The addition of directional information obtained by the timing-based separation of Cherenkov light from scintillation light would significantly enhance background suppression~\cite{direction2014, chessPRC, chessEPJC}. In kiloton-scale detectors, this separation can be improved by using a scintillator with a narrow emission spectrum shifted to shorter wavelengths~\cite{direction2014}. Quantum dot-doped scintillators 
are a good candidate for this application and provide robust chemistry for suspending a variety of isotopes in the scintillator. 

Previous work has focused on scintillators doped with core-shell quantum dots~\cite{qdot1,qdot2}. In this paper, we investigate a promising new class of quantum dots, perovskite nanocrystals. These are also called perovskite nanoplatelets due to their flat geometry. This work builds on Ref.~\cite{perovskite2016}; we verify the absorption and emission spectra of the scintillator cocktails and perform light yield measurements with a $^{22}$Na gamma source. 

\section{Perovskite Nanocrystals}
\label{sec:perosvkites}

\begin{figure}[htbp]
\centering 
\includegraphics[trim=4.5cm 3.5cm 4.5cm 0cm, clip=true, width=0.45 \textwidth]{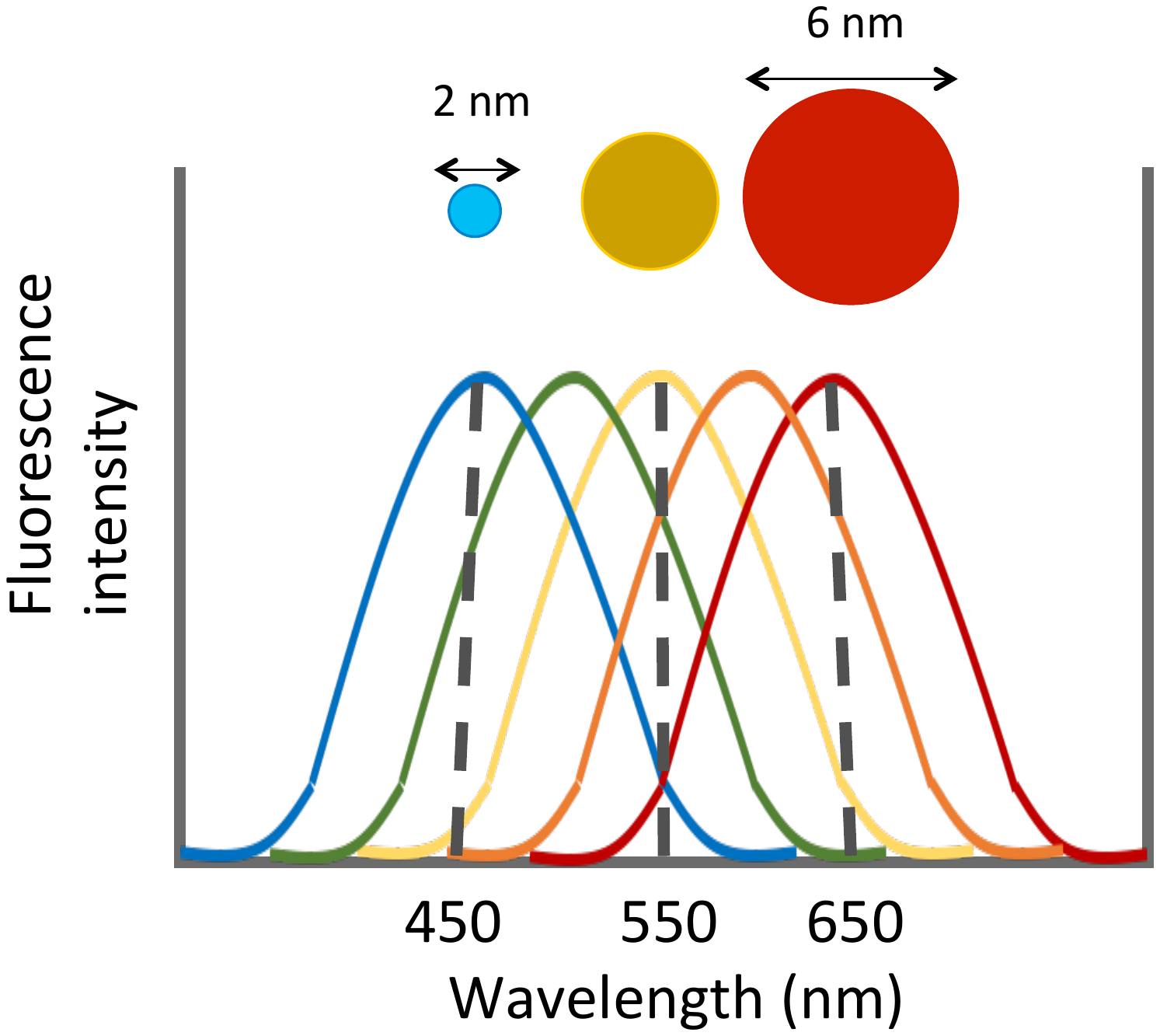}
\includegraphics[trim=2.25cm 2.5cm 4.5cm 3cm, clip=true, width=0.45 \textwidth]{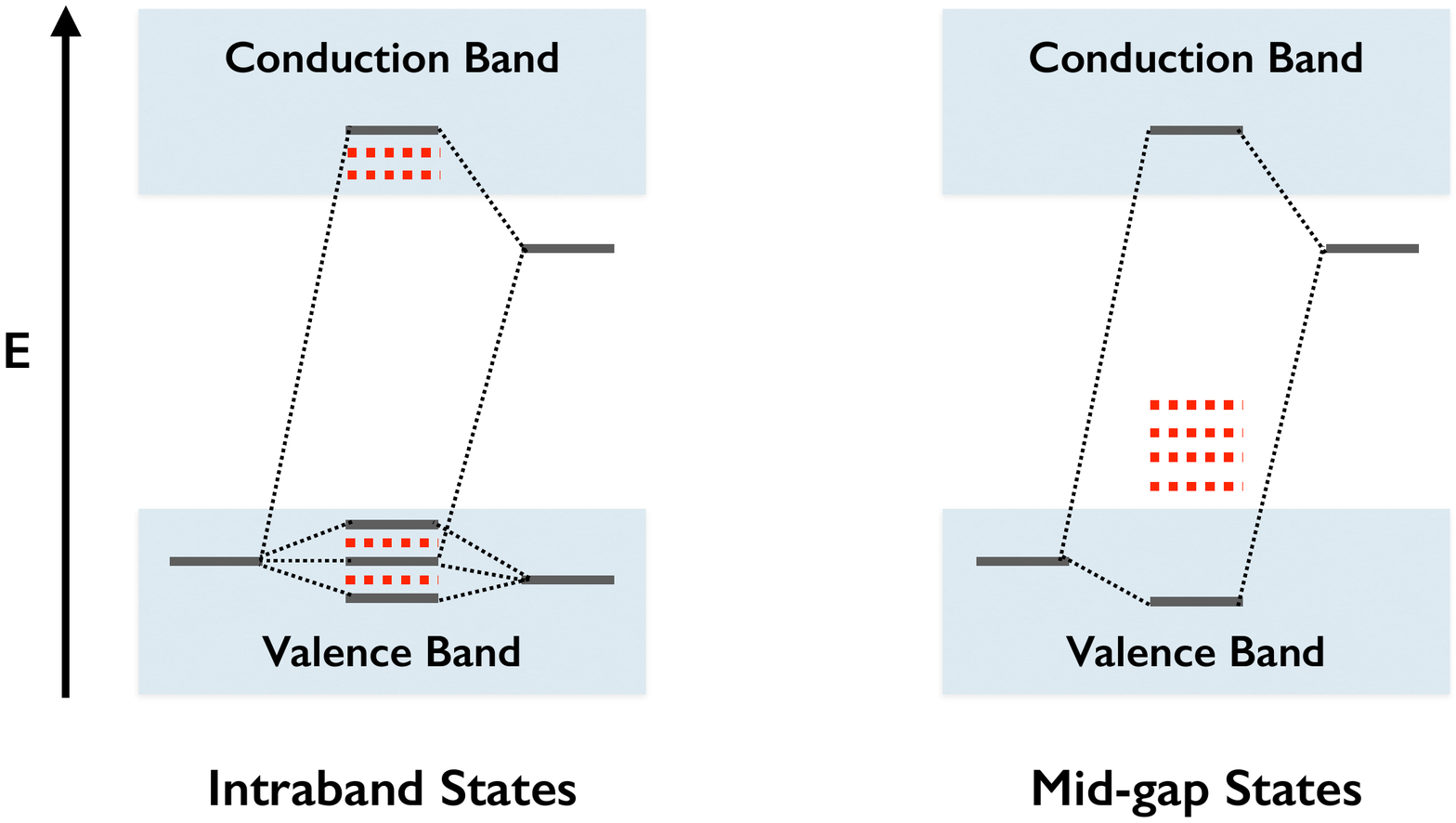}
\caption{\label{qdotIntro} \textit{Left:} Increasing the diameter of the nanocrystal increases the emission wavelength. \textit{Right:} Molecular orbital diagrams adapted from Ref.~\cite{perovskites_Sun2018}. In the intraband example, defect states (red dashes) are contained in either the valence or conduction bands. Moreover, these states are shallow and transient. In the Mid-gap example, defect states form mid-gap, and are not as shallow, decaying non-radiatively, thus reducing the light output.}
\end{figure}

Quantum dots are semiconducting crystals with sizes on the order of nanometers. In this regime, the material's mobile charge carriers are confined to an effective box. This quantum confinement causes the crystal's band gap to increase with decreasing crystal size, and thus blue-shifts its photoluminescence (see Figure~\ref{qdotIntro} \textit{Left}). The tunable, size-dependent optical properties of quantum dots have applications in photocatalysis and solar cells, as well as in light emitting devices including televisions and LEDs. 

The success of these applications requires highly luminescent crystals. Crystal defects such as dangling surface bonds and vacancies often appear in quantum dots. These defects create transition states within the bandgap (see Figure~\ref{qdotIntro} \textit{Right}). This allows non-radiative recombination, thus reducing the light output. To combat these defects, the canonical core quantum dot can be encased in a protective shell made of another nanocrystal. This encasing can be chosen to have the larger bandgap relative to the core crystal in order to shield the core from the formation of defects. Although the addition of a shell has been shown to increase the stability and luminescence of core quantum dots, these shelled dots are often difficult to fabricate, requiring high temperatures and inert atmospheres. Moreover, these procedures are difficult to control and the resulting core-shell quantum dots size often varies from batch to batch in both the extent of passivation and in the thickness of the shell.

\begin{figure}[t]
\centering 
\includegraphics[trim=0cm 3.0cm 0cm 0cm, clip=true, width=0.9\textwidth]{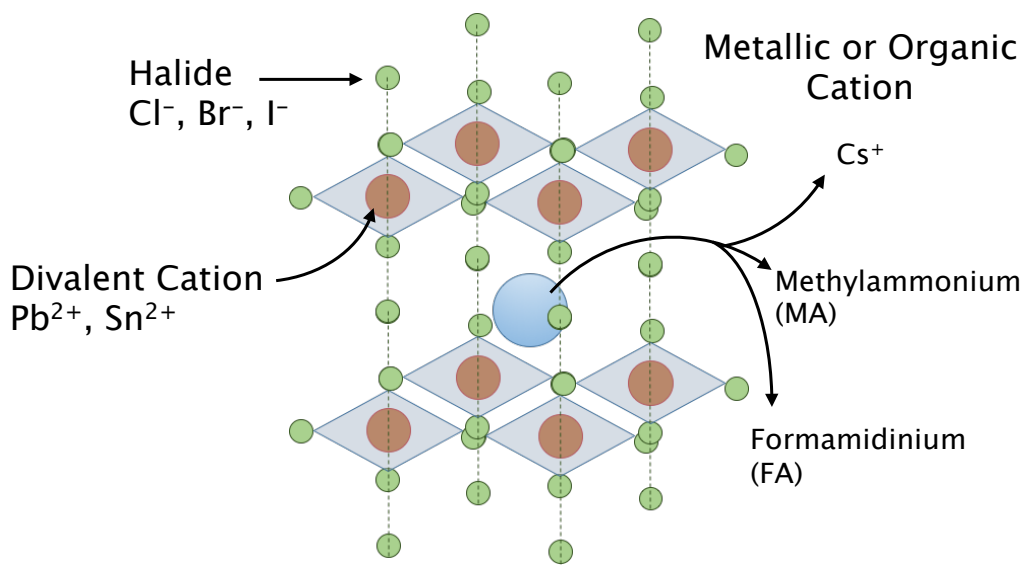}
\caption{\label{qdotJustCrystal} \textit{Right:} Perovskite nanocrystals have the structure ABX$_3$, where A is a monovalent organic or inorganic cation, B is a divalent metal cation, and X is a halide anion. In this work, we study lead\,(II) crystals with bromide and iodide halides and cesium, methylammonium, and formamidinium cations.}
\end{figure}

A new type of quantum dot with a perovskite structure has emerged as a promising material, notable for its high luminescence and ease of synthesis. The bulk perovskite crystal structure has the formula ABX$_3$, where A is a monovalent organic or inorganic cation, B is a divalent metal cation, and X is a halide anion (see Figure~\ref{qdotJustCrystal}). The addition of ligands (L) in the synthesis prevents the formation of of the bulk perovskite, leaving nanoplatelets with thickness dependent on the starting ratio of LX:BX$_2$:AX . Changing the A, B, or X components of this crystal to smaller or larger atoms moves the emission wavelength to shorter or longer wavelengths. Moreover, these dots can be made by solvent precipitation, a more reproducible technique than hot-injection.

Perovskite nanocrystals are unique in their tolerance to defects on the crystal surface, even without passivation by a second nanomaterial. This property is due to the material's electronic properties and ionic bonding character, which cause defect states to form in either the valence or conduction band, rather than in the band gap. These intraband states come from the antibonding character of the valence maximum. This antibonding character causes nonbonding defect states to form below this maximum, and thus in the valence band. In the conduction band, the minimum is stable due to strong spin-orbit coupling, causing only shallow defect states to form. To first order, these non-bonding states are contained in the conduction band, leaving a clean bandgap. By contrast, defect-intolerant nanomaterials exhibit a valence band maximum with bonding character, causing defect states to form mid-bandgap. Without strong spin-orbit coupling from the conduction band, these states tend to be less shallow, and thus less transient than intraband states. These mid-bandgap states undergo non-radiative recombination, decreasing the light output of the crystal.

The most-studied crystal structure is that of the lead-based perovskite nanocrystals, B\,$=$\,Pb\,(II) and is therefore the focus of this work. The search for alternatives to lead that can be used in perovskites must consider the materials' electronic properties. The antibonding character of the valence maximum and strong spin-orbit coupling of the conduction band require two electrons in the outermost \textit{s} orbital. Antimony, bismuth, and tin fit this description, and are active areas of perovskite research. Other more exotic atoms may also be possible.

\section{Perovskite Nanocrystal Synthesis} 

We synthesize perovskite nanocrystals using the method described in Ref.\,\cite{perovskite2016}. Our solvents are purchased from Sigma Aldrich and our perovskite precursors from Great Solar Cell. We use them without further purification or drying. All syntheses are carried out in ambient lab conditions. 

\begin{figure}[htbp]
\centering 
\includegraphics[trim=0cm 4.5cm 0cm 4cm, clip=true, width=0.9\textwidth]{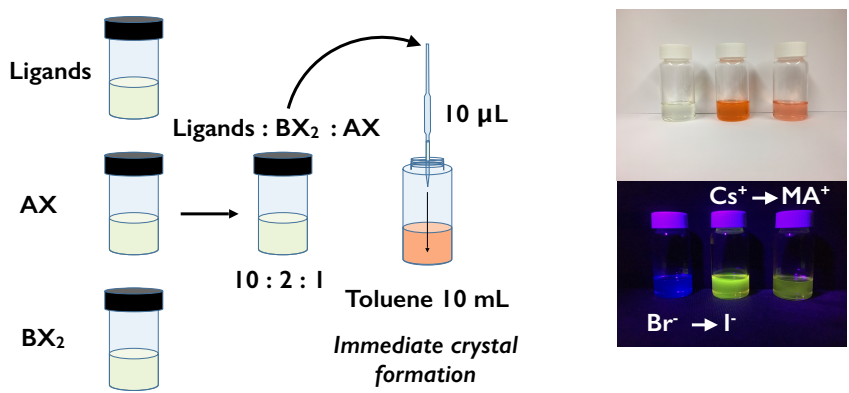}
\caption{\label{qdotSynthesis} \textit{Left:} A diagram of the synthesis, starting with precursors purchased from Great Solar Cell. \textit{Right:} Example samples in ambient light and under UV illumination.}
\end{figure}

The synthesis is summarized in Figure~\ref{qdotSynthesis} (\textit{left}). We first prepare a solution of the perovskite ligands by dissolving an n-butylammonium halide and n-octylammonium halide in dimethylformamide (DMF), each to a concentration of 0.5\,M. Two more solutions, one of halogenated A-site cation (AX)  and one of halogenated B-site cation (BX$_2$) are prepared, each at a concentration of 0.1 M in DMF.   The final precursor solution consisting of a 10:2:1 molar ratio of the ligands, BX$_2$, and AX  is mixed. After pipetting 10\,$\mu$L of this precursor solution into 10\,mL of toluene, perovskite nanoplatelets form immediately, as evidenced by photoluminescence upon radiation with UV light. This is shown in Figure~\ref{qdotSynthesis} (\textit{right}). The synthesis results in a scintillator cocktail with 3.8\,mg/L of Pb. This is consistent with the nanocrystal acting as a secondary wavelength shifter, rather than as a dopant. Future work will increase concentrations to those relevant to applications such as neutrinoless double-beta decay searches~\cite{loadTe2017}. 

We synthesize perovskites from precursors containing two possible halides (bromide and iodide) and three possible A-site cations (cesium, methylammonium, and formamidinium). Of the six possible combinations, the only one not prepared is the brominated cesium perovskite; CsBr has poor solubility in DMF, and requires mixing with highly toxic dimethylsulfoxide (DMSO). 

\section{Fluorescence and Absorption Measurements}

\begin{figure}[htbp]
\centering  
\includegraphics[width=0.5\textwidth]
{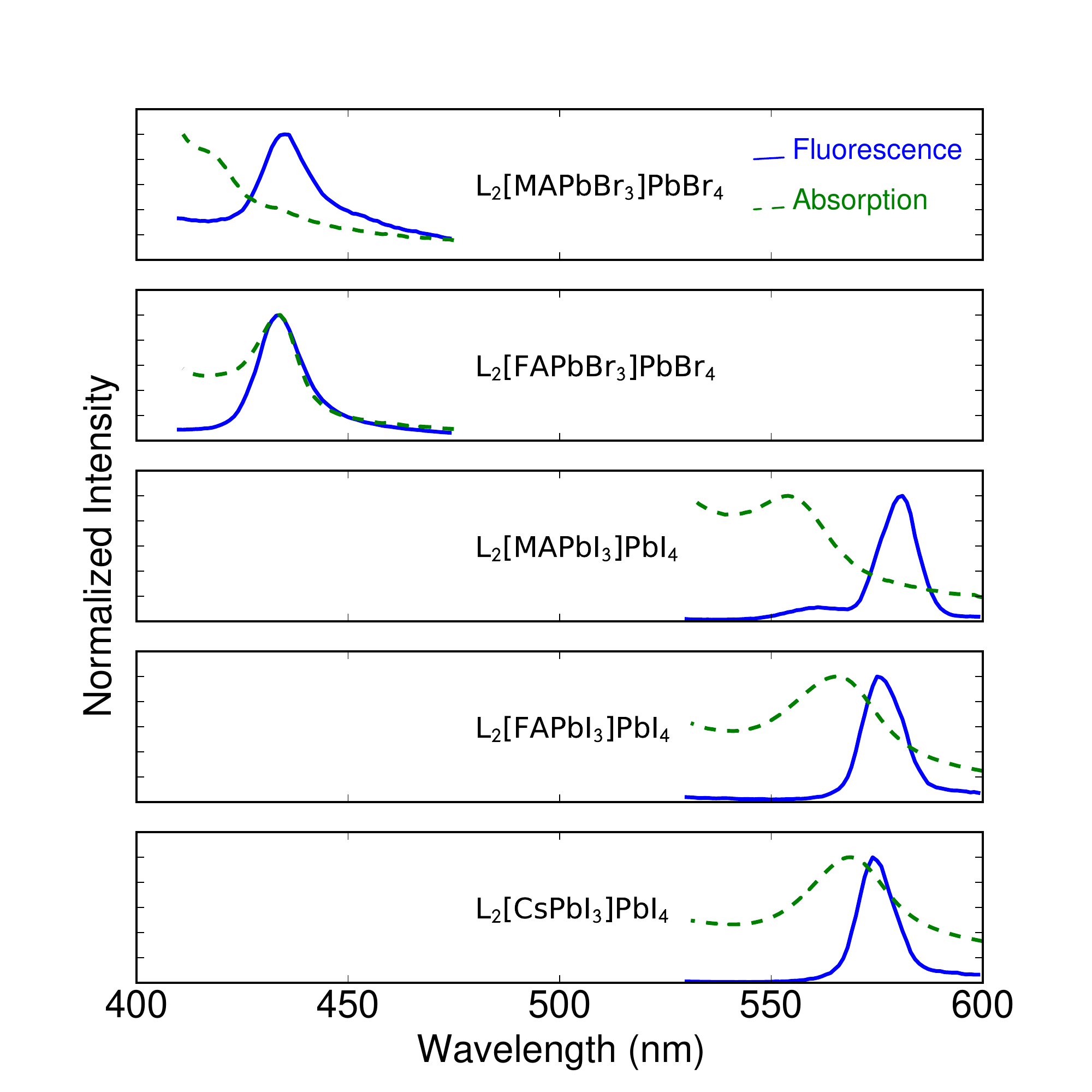}
\caption{\label{qdotEmissionAndAbsorption} The absorption and fluorescence emission spectra for the prepared samples.}
\end{figure}

In order to verify the success of our synthesis and to understand the optical properties of the perovskite nanocrystal and toluene cocktails, fluorescence and absorption measurements are made. We use a Carey Eclipse fluorimeter and a Carey 5000 UVvis spectrophotometer that are available through the Eni-MIT Solar Frontiers Shared Experimental Facilities. The results of these measurements are shown in Figure~\ref{qdotEmissionAndAbsorption}.

The peak emission and absorption values are tabulated in Table~\ref{tab1} and compared to those obtained in Ref.~\cite{perovskite2016}. The peak emission value is well defined. For these nanocrystals, the absorption should have a small peak at wavelengths slightly shorter than the peak emission and then a rise at even smaller wavelengths. This small peak before the rise is consistent with what was observed for absorption measurements. The values agree with the exception of the emission of \CsPbI and the absorption of \MAPbBr. We observe a broader \MAPbBr spectrum, so the absorption value is less well defined, although the general features of the spectrum are correct. The longer wavelength of the \CsPbI emission indicates that the nanocrystal is larger than intended. The precursors for this nanocrystal were slightly older, and it is possible that the precursor was not mixed sufficiently. This serves as useful experience for future studies.

\begin{table}[htbp]
\centering
\caption{\label{tab1} Results of fluorescence and absorption measurements.}
\smallskip
\begin{tabular}{l c c c c}
\hline
 & \multicolumn{2}{c}{Absorbance (nm)}&\multicolumn{2}{c}{Fluorescence (nm)} \\
Sample & Ref.~\cite{perovskite2016} & This work & Ref.~\cite{perovskite2016} & This work \\
\hline
\MAPbBr & 431 & 415 & 437.3 & 435 \\
\FAPbBr & 434 & 433 & 439 & 433 \\
\CsPbI  & 553 & 554 & 561.1 & 581 \\
\MAPbI & 566 & 566 & 573.9 & 576 \\
\FAPbI & 566 & 569 & 575 & 573 \\
\hline
\end{tabular}
\end{table}

\section{Light-yield measurements}

\begin{figure}[htbp]
\centering  
\includegraphics[width=0.45\textwidth]{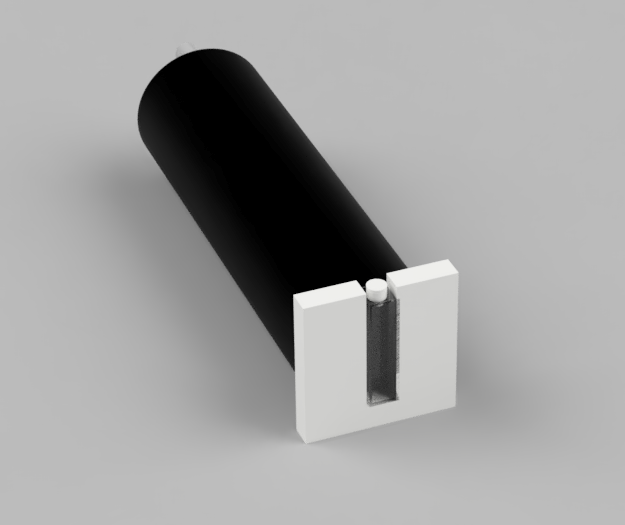}
\includegraphics[width=0.45\textwidth]{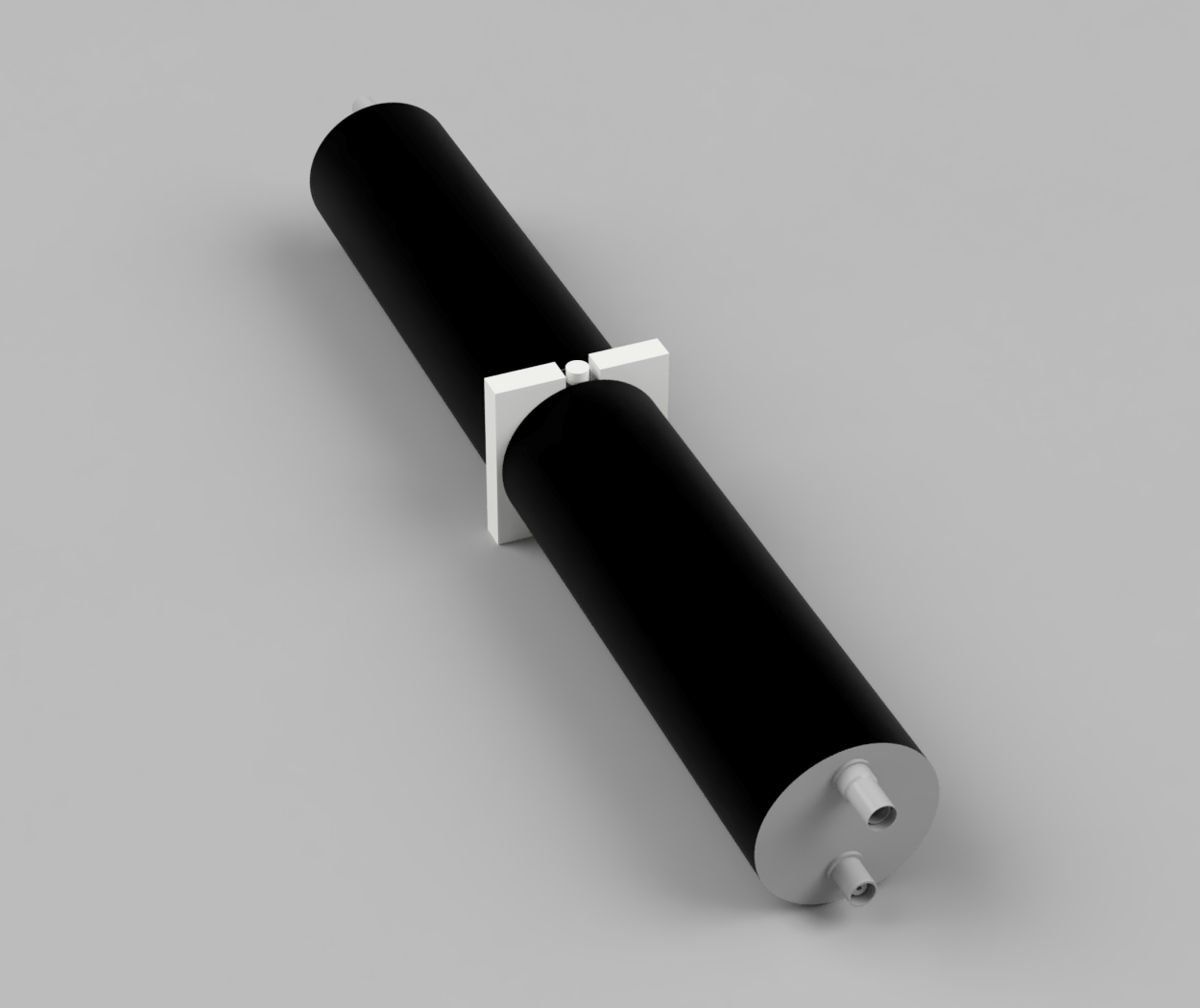}
\caption{\label{LightYieldApparatus} The light yield measurement stand  consists of two PMTs coupled to a quartz cuvette with optical grease. A Teflon holder was fabricated to hold the cuvette. The cuvette holder and PMTs were held together with black tape (not shown) to limit reflections, and a radioactive source (not shown) is taped to the top of the cuvette holder.}
\end{figure}

The light yield of the scintillator cocktails is measured with an apparatus containing two photomultiplier tubes (PMTs) and a quartz cuvette, arranged as shown in Figure \ref{LightYieldApparatus}. The PMTs are Hamamatsu model R13089, which is a high quantum efficiency 2-inch PMT. The PMTs are operated at 1750\,V, provided by a Wiener EDS 30330n ISEG MPOD high voltage module. The cuvette used is a Spectrosil 21-Q-10, with exterior dimensions of 12.5\,mm $\times$ 12.5\,mm $\times$ 4.8\,mm. 

The signal from each PMT is divided, with half of it read out directly by the digitizer, a CAEN V1742 digitizer sampling at 5\,GS/s. The other half of each signal is amplified by a factor of 10 in an MMIC fixed-gain low-noise amplifier and used to produce a 5\,ns-wide NIM logic gate; the coincidence of these gates, produced by a CAEN V976 FIFO logic unit, is used to trigger the digitization of the un-amplified signals. Each recorded waveform has 1024 samples and is written to disk in HDF-5 by the data acquisition computer for offline processing.

For each acquired waveform, a trigger point is obtained by a level discriminator set to -10\,mV. The total amount of anode charge is then determined by integrating the waveform from -10\,ns to 80\,ns, relative to the trigger point, avoiding a cable reflection which begins at 90 ns. The charge in ADC units is then converted to pC by dividing by the nominal 50\,$\Omega$ termination of the board. The PMTs are roughly gain matched at their operating voltage.

A $^{22}$Na button source is placed on top of the cuvette and is used as the calibration source. $^{22}$Na is a $\beta^+$ emitter that decays to the first excited state of $^{22}$Ne, producing a 1.275\,MeV $\gamma$ from the de-excitation of $^{22}$Ne and two 0.511\,MeV $\gamma$'s from the annihilation of the $\beta^+$. This allows us to measure the light yield at two energies simultaneously.   

In order to extract the light yield of the scintillator cocktail, a simple simulation is used to model the spectral shape. The cuvette and scintillator are modeled in a Geant4 simulation~\cite{geant4one,geant4two}.  Both $\gamma$'s are simulated and the total energy deposited by the scattering electrons is recorded. The result is that the Compton edge is smeared to lower energies due to the electrons hitting the wall of the cuvette. This result is then convolved with a Gaussian detector response function and used to fit the data. The fit has three variables: the normalization, a linear calibration constant (to convert from charge to energy) and the width ($\sigma$) of the Gaussian energy resolution. 

\begin{figure}[htbp]
\centering  
\includegraphics[width=0.45\textwidth]{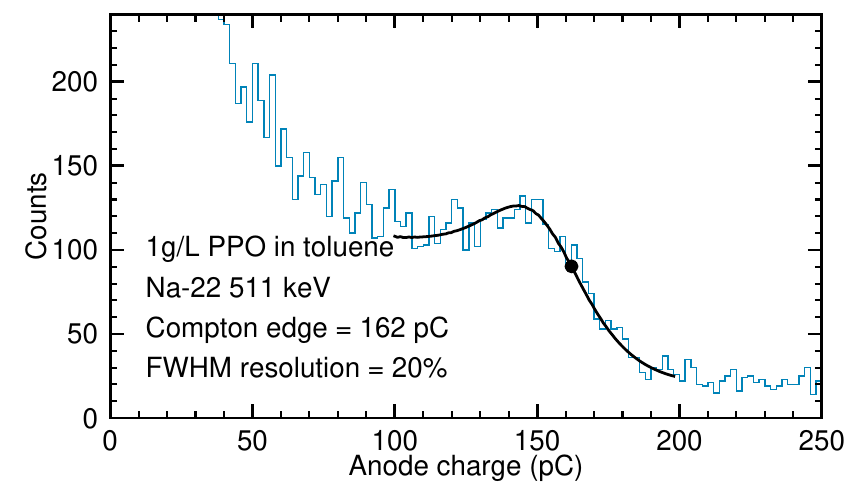}
\includegraphics[width=0.45\textwidth]{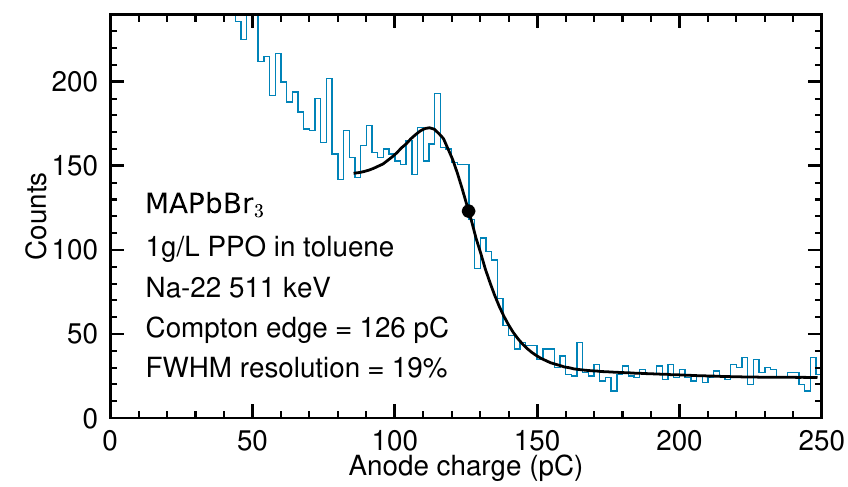}
\caption{\label{Spectrum} Sample energy spectrum, (Left) Toluene with 1\,g/L PPO and (Right) the \MAPbBr cocktail, taken with the light yield apparatus. The spectral shape is modeled using a Geant4 simulation and fit to the spectrum with three free parameters to extract the light yield.}
\end{figure}

\begin{table}[htbp]
\centering
\caption{\label{tab2} The light yield results are presented both in terms of the total charge collected and relative to the light yield of the 511\,keV $\gamma$ in toluene with 1\,g/L PPO. The relative light yield is also given after correction for the PMT quantum efficiency at the peak emission wavelength.}
\smallskip
\begin{tabular}{c c c c c c}
\hline
Sample & Emission [nm] & Light Yield [pC] & Ratio & PMT QE & Corrected Ratio \\
\hline
Toluene with 1\,g/L PPO & 380 & 162 & 1.00 & 0.350 & 1.00 \\
\MAPbBr & 435 & 126 & 0.78 & 0.318 & 0.86 \\
\FAPbBr & 433 & 124 & 0.77 & 0.316 & 0.85 \\
\CsPbI  & 582 & 50 & 0.31 & 0.063 & 1.71 \\
\MAPbI  & 576 & 48 & 0.30 & 0.070 & 1.48 \\
\FAPbI  & 573 & 47 & 0.29 & 0.074 & 1.37 \\
\hline
\end{tabular}
\end{table}

The scintillating cocktails are made from toluene with 1\,g/L 2,5-diphenyloxazole (PPO) and one of the five nanocrystal samples. Light yield measurements are taken for each sample and for a benchmark cocktail, toluene with 1\,g/L PPO. The results of these fits are summarized in Table~\ref{tab2}. We benchmark against the light yield of toluene with 1\,g/L PPO at 511\,keV. We present both the raw numbers and those relative to the benchmark cocktail. 

As expected, the addition of the heavy metal Pb to the scintillator reduces the light yield at $\sim$400\,nm. This is the range where the both the toluene emission spectrum and PMT quantum efficiency (QE) have their peaks. The current Pb concentration is low; future work will explore how this reduction changes with loading fraction.    

It is not surprising that the long-wavelength-emitting perovskites are significantly dimmer than the benchmark cocktail, since the PMT QE is severely reduced at longer wavelengths. If a correction is applied for the QE, however, these scintillators are significantly brighter than traditional scintillators. This is an interesting result which motivates future study of perovskite nanocrystals for this application.

\section{Conclusion}
In this work, we successfully synthesize and measure the light yield of perovskite nanocrystal-doped liquid scintillator. The synthesis we use makes dilute samples. However, the nanocrystal additive still dominates the response of the scintillator. The cocktails in the $\sim$400\,nm range show 15-20\% reduction in light yield. Intriguingly, the cocktails at longer wavelengths show increases in light yield on the order of 50\%. 

The light yield is a combination of the intrinsic bandgap of the crystal and the ligand chemistry. These longer wavelength nanocrystals are most relevant to device applications, therefore they have been studied more extensively than other varieties. Tuning the ligand chemistry and using different perovskite crystals would both increase the light yield and allow for different isotope dopants. This will be the focus of future work, in addition to studies of nanocrystal concentrations and stability.

\acknowledgments
This work is supported by NSF award number 1554875. We thank Anna Osherov for training and support in the fluorimeter and UVvis measurements and we thank Watcharaphol Paritmongkol for advice on the synthesis. The authors thank Riccardo Comin, Jake Siegel and William Tisdale for helpful discussions on perovskite nanocrystals. This work is part of the NuDot experiment and we thank that collaboration for many useful discussions.

\bibliographystyle{JHEP}
\bibliography{bibliography_scint_short_2018}

\providecommand{\href}[2]{#2}\begingroup\raggedright\begin{thebibliography}{10}

\bibitem{kamReactor2013}
{\scshape KamLAND} collaboration, A.~Gando et~al., \emph{{Reactor On-Off
  Antineutrino Measurement with KamLAND}},
  \href{http://dx.doi.org/10.1103/PhysRevD.88.033001}{\emph{Phys. Rev.}
  {\bfseries D88} (2013) 033001},
  [\href{https://arxiv.org/abs/1303.4667}{{\ttfamily 1303.4667}}].

\bibitem{borexinoSoloar2011}
G.~Bellini et~al., \emph{{Precision measurement of the 7Be solar neutrino
  interaction rate in Borexino}},
  \href{http://dx.doi.org/10.1103/PhysRevLett.107.141302}{\emph{Phys. Rev.
  Lett.} {\bfseries 107} (2011) 141302},
  [\href{https://arxiv.org/abs/1104.1816}{{\ttfamily 1104.1816}}].

\bibitem{dayabay2016}
{\scshape Daya Bay} collaboration, F.~P. An et~al., \emph{{Measurement of
  electron antineutrino oscillation based on 1230 days of operation of the Daya
  Bay experiment}},
  \href{http://dx.doi.org/10.1103/PhysRevD.95.072006}{\emph{Phys. Rev.}
  {\bfseries D95} (2017) 072006},
  [\href{https://arxiv.org/abs/1610.04802}{{\ttfamily 1610.04802}}].

\bibitem{reno2018s}
{\scshape RENO} collaboration, S.~H. Seo et~al., \emph{{Spectral Measurement of
  the Electron Antineutrino Oscillation Amplitude and Frequency using 500 Live
  Days of RENO Data}},
  \href{http://dx.doi.org/10.1103/PhysRevD.98.012002}{\emph{Phys. Rev.}
  {\bfseries D98} (2018) 012002},
  [\href{https://arxiv.org/abs/1610.04326}{{\ttfamily 1610.04326}}].

\bibitem{doublechooz2014}
{\scshape Double Chooz} collaboration, Y.~Abe et~al., \emph{{Improved
  measurements of the neutrino mixing angle $\theta_{13}$ with the Double Chooz
  detector}}, \href{http://dx.doi.org/10.1007/JHEP02(2015)074,
  10.1007/JHEP10(2014)086}{\emph{JHEP} {\bfseries 10} (2014) 086},
  [\href{https://arxiv.org/abs/1406.7763}{{\ttfamily 1406.7763}}].

\bibitem{prospect2018}
{\scshape PROSPECT} collaboration, J.~Ashenfelter et~al., \emph{{First search
  for short-baseline neutrino oscillations at HFIR with PROSPECT}},
  \href{https://arxiv.org/abs/1806.02784}{{\ttfamily 1806.02784}}.

\bibitem{direction2014}
C.~Aberle, A.~Elagin, H.~J. Frisch, M.~Wetstein and L.~Winslow,
  \emph{{Measuring Directionality in Double-Beta Decay and Neutrino
  Interactions with Kiloton-Scale Scintillation Detectors}},
  \href{http://dx.doi.org/10.1088/1748-0221/9/06/P06012}{\emph{JINST}
  {\bfseries 9} (2014) P06012},
  [\href{https://arxiv.org/abs/1307.5813}{{\ttfamily 1307.5813}}].

\bibitem{chessPRC}
J.~Caravaca, F.~B. Descamps, B.~J. Land, J.~Wallig, M.~Yeh and G.~D.
  Orebi~Gann, \emph{{Experiment to demonstrate separation of Cherenkov and
  scintillation signals}},
  \href{http://dx.doi.org/10.1103/PhysRevC.95.055801}{\emph{Phys. Rev.}
  {\bfseries C95} (2017) 055801},
  [\href{https://arxiv.org/abs/1610.02029}{{\ttfamily 1610.02029}}].

\bibitem{chessEPJC}
J.~Caravaca, F.~B. Descamps, B.~J. Land, M.~Yeh and G.~D. Orebi~Gann,
  \emph{{Cherenkov and Scintillation Light Separation in Organic Liquid
  Scintillators}},
  \href{http://dx.doi.org/10.1140/epjc/s10052-017-5380-x}{\emph{Eur. Phys. J.}
  {\bfseries C77} (2017) 811},
  [\href{https://arxiv.org/abs/1610.02011}{{\ttfamily 1610.02011}}].

\bibitem{qdot1}
L.~Winslow and R.~Simpson, \emph{{Characterizing Quantum-Dot-Doped Liquid
  Scintillator for Applications to Neutrino Detectors}},
  \href{http://dx.doi.org/10.1088/1748-0221/7/07/P07010}{\emph{JINST}
  {\bfseries 7} (2012) P07010},
  [\href{https://arxiv.org/abs/1202.4733}{{\ttfamily 1202.4733}}].

\bibitem{qdot2}
C.~Aberle, J.~J. Li, S.~Weiss and L.~Winslow, \emph{{Optical Properties of
  Quantum-Dot-Doped Liquid Scintillators}},
  \href{http://dx.doi.org/10.1088/1748-0221/8/10/P10015}{\emph{JINST}
  {\bfseries 8} (2013) P10015},
  [\href{https://arxiv.org/abs/1307.4742}{{\ttfamily 1307.4742}}].

\bibitem{perovskite2016}
M.~C. Weidman, M.~Seitz, S.~D. Stranks and W.~A. Tisdale, \emph{Highly tunable
  colloidal perovskite nanoplatelets through variable cation, metal, and halide
  composition}, \href{http://dx.doi.org/10.1021/acsnano.6b03496}{\emph{ACS
  Nano} {\bfseries 10} (2016) 7830--7839},
  [\href{https://arxiv.org/abs/https://doi.org/10.1021/acsnano.6b03496}{{\ttfamily
  https://doi.org/10.1021/acsnano.6b03496}}].

\bibitem{perovskites_Sun2018}
J.~Sun, J.~Yang, J.~I. Lee, J.~H. Cho and M.~S. Kang, \emph{Lead-free
  perovskite nanocrystals for light-emitting devices},
  \href{http://dx.doi.org/10.1021/acs.jpclett.8b00301}{\emph{The Journal of
  Physical Chemistry Letters} {\bfseries 9} (2018) 1573--1583},
  [\href{https://arxiv.org/abs/https://doi.org/10.1021/acs.jpclett.8b00301}{{\ttfamily
  https://doi.org/10.1021/acs.jpclett.8b00301}}].

\bibitem{loadTe2017}
S.~Biller and S.~Manecki, \emph{{A New Technique to Load 130Te in Liquid
  Scintillator for Neutrinoless Double Beta Decay Experiments}}, {\emph{Journal
  of Physics: Conference Series} {\bfseries 888} (2017) 012084}.

\bibitem{geant4one}
{\scshape GEANT4} collaboration, S.~Agostinelli et~al., \emph{{GEANT4: A
  Simulation toolkit}},
  \href{http://dx.doi.org/10.1016/S0168-9002(03)01368-8}{\emph{Nucl.Instrum.Meth.}
  {\bfseries A506} (2003) 250--303}.

\bibitem{geant4two}
{\scshape GEANT4} collaboration, J.~Allison et~al., \emph{{Geant4 developments
  and applications}},
  \href{http://dx.doi.org/10.1109/TNS.2006.869826}{\emph{Nuclear Science, IEEE
  Transactions on} {\bfseries 53} (2006) 270--278}.

\end{thebibliography}\endgroup

\end{document}